\documentclass[prl,10pt,twocolumn,showpacs,amssymb,amsfonts,amsmath,groupedaddress,floatfix,nofootinbib]{revtex4}

\usepackage{latexsym,graphicx,graphics} % epsf,psfrag

\begin{document}

%\ \\[-1.2cm]
\title{Fair and optimistic quantum contract signing}

%\ \\[-0.7cm]
\author{N. Paunkovi\'c$^{1}$,  J. Bouda$^{2}$ and P. Mateus$^{1}$}
\affiliation{$^{1}$SQIG -- Instituto de Telecomunica\c{c}\~oes, IST, TULisbon, Av. Rovisco Pais 1049-001 Lisbon, Portugal}
\affiliation{$^{2}$Faculty of Informatics, Masaryk University, Botanick\'a 68a, 60200, Brno, Czech Republic}

%\date{\today}{\today}

\begin{abstract}
%\ \\[-0.5cm]
We present a fair and optimistic \cite{ben:90,aso:97} quantum contract signing protocol between two clients that requires no communication with the third trusted party during the exchange phase. We discuss its fairness and show that it is possible to design such a protocol for which the probability of a dishonest client to cheat becomes negligible, and scales as $N^{-1/2}$, where $N$ is the number of messages exchanged between the clients. Our protocol is not based on the exchange of {\em signed} messages: its fairness is based on the laws of quantum mechanics. Thus, it is abuse-free \cite{abuse-free_1}, and the clients do not have to generate new keys for each message during the Exchange phase. We discuss a real-life scenario when the measurement errors and qubit state corruption due to noisy channels occur and argue that for real, good enough measurement apparatus and transmission channels, our protocol would still be fair. Our protocol could be implemented by today's technology, as it requires in essence the same type of apparatus as the one needed for BB84 cryptographic protocol \cite{bb84}. Finally, we briefly discuss two alternative versions of the protocol, one that uses only two states (based on B92 protocol \cite{b92}) and the other that uses entangled pairs (based on \cite{ekert:91}), and show that it is possible to generalize our protocol to an arbitrary number of clients.
%\ \\[-0.5cm]
\end{abstract}

%\ \\[-0.8cm]

\maketitle

%%%%%%%%%%%%%%%%%%%%%%%%%%%%%%%%%%%%%%%%%%%%%%%%%%%%%%%%%%%%%%%%%%%%%%%%%%%%%%%%%%%%%%%%%%%%%%%%%%%%%%%%%%%%%%%%%%%%%%%
\section{I. Introduction}
\label{sec:introduction}
%%%%%%%%%%%%%%%%%%%%%%%%%%%%%%%%%%%%%%%%%%%%%%%%%%%%%%%%%%%%%%%%%%%%%%%%%%%%%%%%%%%%%%%%%%%%%%%%%%%%%%%%%%%%%%%%%%%%%%%

Contract signing \cite{Even.Yacobi:Relationsamongpublic-1980} is an important security task with many applications, namely to stock market and others \cite{CSapplications}. It is a two party protocol between Alice and Bob who share a common contract and want to exchange each others' commitments to it, thus binding to the terms of the contract. Usually, commitment is done by signing the contract on the spot. 

With the technology development, situations when parties involved are physically far apart become more relevant every day -- distant people can communicate using ordinary or e-mail, internet, etc. This poses new challenges to the problem. Forcing spatially distant parties to exchange signatures opens the possibility of a fraud: Bob may get the commitment from Alice (a copy of the contract with her signature on it) without committing himself, which creates an {\em unfair situation}. Indeed, having Alice's commitment enables Bob to appeal to a judge to bind the contract (i.e., to enforce; declare it valid), by showing Alice's commitment to the contract (together with his signature stamped on it). On the other hand, although Alice did commit, she cannot prove it (prove that she sent her commitment to Bob) and thus cannot appeal to a judge. Moreover, she cannot prove that she did not receive Bob's commitment, so he can safely claim that he behaved honestly and sent his commitment to Alice. Note that initially Bob did not commit, but having Alice's commitment puts him in a position to {\em later in time} choose whether to sign the contract or not, and thus bind it or not, while Alice has no power to do either of the two. 

This situation is particularly relevant in a stock market, for example, where prices of stocks may range over time, but agents must commit {\em beforehand} to sell/buy at a certain time in the {\em future}, for previously fixed prices. The unfairness allows Bob to make risky decisions in the stock market without being bound to them, unless he decides so. The problem when distant parties wish to commit to a common contract lies in the impossibility for an agent, say Alice, to prove whether she has indeed committed to it or not. Often, the contract signing problem is said to be a variant of the so-called {\em return receipt} (or {\em certified mail}) {\em problem}, in which parties involved exchange mails between each other asking for the proof confirming whether the other side received the message, or not. 

A simple solution to this unfair situation is to have a trusted third party (usually referred to as Trent) mediating the transaction - Alice and Bob send their commitments to Trent, who then returns the receipts to the senders, and performs the message exchange {\it only} upon receiving both of the commitments. However, Trent's time and resources are expensive and should be avoided as much as possible. Unfortunately, it has been shown that there is no fair and viable contract signing protocol \cite{Even.Yacobi:Relationsamongpublic-1980,fis:lyn:pat:85}, unless during the signature exchange phase the signing parties communicate with a common trusted agent, i.e., Trent. By {\em fair} protocol we mean that either both parties get each other's commitment or none gets. By {\em viable} protocol we mean that, if both parties behave honestly, they will both get each others' commitments.

The essence of the proof of the above impossibility result is rather simple, and is related with the impossibility of establishing distributed consensus in asynchronous networks \cite{fis:lyn:pat:85}. The simple assumption we need for the proof is the following: the protocol consists of a number of messages exchanged between the two parties, so that eventually, upon the termination of the protocol, both Alice and Bob acquire the signature of the other. This can be done by either sending pieces of signatures in each message, or in a more sophisticated scenarios, by sending partial information needed to calculate, upon running a complex algorithm, the signature needed \cite{even:82,even:goldreich:lempel:85}. We see that if such protocol existed, it would have a final step where one message is exchanged, say, from Bob to Alice. In that case, before sending his last message, Bob would already have all the information required for him to compute Alice's signature of the contract (and Alice does not). Therefore, if he does not send the last message, the protocol reaches an unfair state. Note the essential importance of the asynchronousity -- it is not possible for {\em distant} parties to arrange in advance that messages are sent {\em simultaneously}. 

One way to come around this difficulty is to consider {\em optimistic} protocols that do not require communication with Trent unless something wrong comes up (some message is missing, etc.) \cite{aso:97}. In such protocols, the clients contact Trent regarding the given contract before the actual signing takes place. Trent notifies the request and assigns the particular contract with the clients, this way {\em initializing} the signing protocol. After that, the clients {\em exchange} messages between each other such that, if the protocol is executed correctly by both sides, both will end up with signed messages. In case something goes wrong (message that is not according to the protocol is sent, or communication interrupted), Trent is contacted in order to {\em bind} the contract.

%Optimistic protocols exist for which the fairness condition is relaxed probabilistically. 
Another workaround is to relax the fairness condition probabilistically. {\it Probabilistic fairness} allows one agent to have at most $\varepsilon$ more probability of binding the contract over the other agent, at each step of the protocol. In this case, for an arbitrarily small $\varepsilon$, classical solutions have been found where the number of exchanged messages between the agents is minimized \cite{rabin:83,ben:90}. In addition to being (probabilistically) fair, in the protocol by Rabin \cite{rabin:83} the joint probability that one agent can bind the contract, while the other cannot, is also always smaller than given $\varepsilon$. Moreover, the second protocol by Ben-Or {\em et. al} \cite{ben:90} satisfies even stronger condition: the conditional probability that an agent cannot bind the contract, given that the other can, can be made arbitrarily small. Note that the two notions are not exclusive: the protocol \cite{ben:90} is both fair and optimistic.

In this paper, we present a (probabilistically) {\em fair} contract signing protocol where {\em no information} with a trusted third party (Trent) is exchanged during the exchange phase. This way, it avoids possible communication bottlenecks that are otherwise inherent when involving Trent. Information exchange takes place during the initialization phase and possibly later during the (contract) binding phase (the protocol is {\em optimistic} \cite{aso:97}: Trent is rarely asked to bind the contract due to protocol fairness - cheating does not pay off). Unlike previous classical proposals, in our quantum protocol the messages exchanged between the clients (Alice and Bob) during the exchange phase do {\em not} have to be {\em signed}. Consequently, our protocol is abuse-free \cite{abuse-free_1}: a client has no proof that (s)he communicated with the other client, attempting to sign a {\em given} contract. In our protocol only two signed messages are exchanged. This is very important when one wants to achieve unconditional security. In the case of classical protocols, digital pseudo-signatures \cite{Chaum.Roijakkers-Unconditionally-SecureDigitalSignatures-1991} should be used, where key is one-use and expensive to generate. Finally, our protocol can be used in solving some purely quantum protocols involving timely decisions between spatially distant parties, such was the case of simultaneous dense coding and teleportation \cite{daowen:11}.

In classical cryptography the contract exchange is done in the way that respective participants are learning some information (signed message, etc.) bit by bit, thus increasing their knowledge. In order to bind the contract they have to present the (complete) information to Trent. Our approach is somewhat different, and is based on the laws of quantum physics. 

Quantum systems obey the laws of quantum physics, which exhibit some counterintuitive features that are quite distinct from those of classical physics. The principle of quantum superposition, entanglement and interference, to name just a few, have found numerous applications in the growing field of quantum information and computation, bringing about many advantages of quantum-based information processing protocols over those using classical systems only. Quantum cryptography \cite{bb84,qcryptography-security} guarantees secure communication and devices based on its principles can be already bought on a market today; Shor's algorithm for factoring \cite{Shor} is exponentially faster than any known classical counterpart; the use of entanglement can in some cases of distributed information processing protocols considerably decrease the amount of information exchange between distant parties needed to perform a given task, thus decreasing (quantum) communication complexity of the problem \cite{complexity} and can even eliminate any need of communication that is classically necessary for achieving a goal of common interest of separated parties \cite{pseudo-telepathy} (a feature often referred in literature to as {\em pseudo-telepathy}). For an overview of the field of quantum information and computation, see for example \cite{Nielsen}.

In solving the contract signing problem, we use quantum complementarity which as a consequence has the impossibility to unambiguously discriminate between two quantum states unless they are mutually orthogonal \cite{ivanovic,helstrom}. This fact is at the heart of the security of quantum cryptographic protocols. In particular, as the setups of the famous BB84 quantum key distribution protocol \cite{bb84} and our contract signing protocol have conceptual similarities, so the security of the former is closely related to the fairness of the latter. At the end, we show an alternative version of our contract signing protocol that uses quantum entanglement, a non-local version of complementarity\footnote{Both quantum complementarity and quantum entanglement could be seen as consequences of the Heisenberg uncertainty principle and ultimately, the superposition principle.}. Entanglement provides the possibility of replacing the exchange of classical information with Trent during the exchange phase, by establishing the {\em proper} type of correlations between the distant parties, in a similar fashion as in Ekert's key distribution protocol that uses entangled states \cite{ekert:91}. 

The very same mechanism of commitment to one specific choice can be used to establish e.g., a bit commitment protocol. Note that unconditionally secure bit commitment is not possible without Trent \cite{Mayers-Uncon_secur_quant:1996,Lo+Chau-quant_commi_reall:1997}, although it is realizable using other assumptions as well.
 
The paper is organized as follows. In the next section we present the quantum contract signing protocol that requires no communication with Trent during the exchange phase, is optimistic and fair. In Section III. we present the conditions we require for the protocol to be fair. In the Section IV. we discuss its fairness for the case of ideal measurements restricted to only two alternative single-particle observables. In Section V. we analyze the case of general multi-particle measurements (POVMs) and argue that under the assumption of noisy channels and realistic detectors with error rates, it is still possible to design a fair protocol. We also discuss alternative protocols that, instead of four states, use either two non-orthogonal states, or entangled pairs. Finally, we present a generalization to the case of more than two agents. In Conclusions we present a short overview of the results presented and some of  possible future lines of research.

%We show that it is straightforward to design a protocol in which it is not necessary that both clients are (simultaneously) present during the binding phase. In this scenario, the probability to cheat becomes the joint probability that one client can bind the contract, while the other cannot.

%%%%%%%%%%%%%%%%%%%%%%%%%%%%%%%%%%%%%%%%%%%%%%%%%%%%%%%%%%%%%%%%%%%%%%%%%%%%%%%%%%%%%%%%%%%%%%%%%%%
\section{II. The protocol}
\label{sec:protocol}
%%%%%%%%%%%%%%%%%%%%%%%%%%%%%%%%%%%%%%%%%%%%%%%%%%%%%%%%%%%%%%%%%%%%%%%%%%%%%%%%%%%%%%%%%%%%%%%%%%%

In order for it to be fair, any contract signing protocol has to force a client to make {\it only one} out of two possible choices - accept or reject the contract. A conceptually similar situation occurs in quantum physics, where an observer cannot simultaneously measure two complementary observables, say the position and the velocity of a quantum system. In other words, an observer is forced to choose to measure {\it only one} out of two possible observables and gain information about only one out of two physical properties of the system observed. For example, by measuring the exact position of a quantum particle, we are left with complete uncertainty of its velocity, and vice versa (Heisenberg uncertainty relations).

This basic feature of quantum physics is the essential ingredient of our protocol: instead of sending to Trent the information that explicitly states the acceptance or rejection of the contract, Alice reveals her choice by measuring one of the two complementary observables, thus acquiring information about only one of the two possible features of the system given to her by Trent (and the same for Bob). Gaining information about one feature thus corresponds to the acceptance, while acquiring information about the other corresponds to rejection of the contract. This information can later be used as a proof of client's choice.

As a client's measurement is local, no information is exchanged between a client and Trent, during the exchange phase. Only latter, during the possible binding phase, this (classical) information obtained by a client's measurement is confronted with Trent's (classical) information of the quantum state in which he prepared the quantum system distributed to a client, and thus used as verification of client's choice. 

To ensure the timely decisions, Trent provides Alice with the classical information of the quantum state in which Bob's quantum system is prepared, and vice versa. This way, the clients can confront each others' measurement results with the classical data provided by Trent, thus obtaining each others' commitment choices before a certain fixed moment in time. Since quantum mechanics is essentially a probabilistic theory, the clients are supplied by a number of systems, giving rise to the probabilistic fairness of the protocol.

In our protocol, we use the simplest two-dimensional quantum systems called qubits. The complementary observables could be seen as spin components (for electrons), or linear polarizations (for photons), along two mutually orthogonal axes. We will denote the two observables measured on single qubits as {\it the Accept} observable $\hat{A}$ and {\it the Reject} observable $\hat{R}$. Measuring $\hat{A}$ corresponds to the acceptance, while measuring $\hat{R}$ corresponds to the rejection of the contract. The two observables $\hat{A}$ and $\hat{R}$ are required to be mutually complementary and are given by mutually unbiased bases  \cite{ivanovic} $\mathcal{B}_A = \{ |0\rangle, |1\rangle \}$ ({\it the Accept} basis) and $\mathcal{B}_R = \{ |-\rangle, |+\rangle \}$ ({\it the Reject} basis), respectively, such that $|\pm\rangle=(|1\rangle\pm |0\rangle)/\sqrt{2}$. Both observables have the same eigenvalues, $0$ and $1$, such that
\begin{eqnarray}
\label{observables}
\hat{A} & = & 1\cdot |1\rangle\langle 1| + 0\cdot |0\rangle\langle 0|, \nonumber \\
\hat{R} & = & 1\cdot |+\rangle\langle +| + 0\cdot |-\rangle\langle -|. 
\end{eqnarray}
During the Initialization phase, Trent randomly prepares qubits, each in one of the four states $\{ |0\rangle,|1\rangle,|-\rangle,|+\rangle \}$, taken from the Accept or the Reject basis. Thus, each qubit state $|\psi\rangle \in \{ |0\rangle,|1\rangle,|-\rangle,|+\rangle \}$ is defined by two classical bits $C=(C_b,C_s)$, first of which defines the basis ($C_b=1$ if $|\psi\rangle \in \mathcal{B}_A$, while $C_b=0$ otherwise) while the second defines the particular state from a given basis ($C_s=1$ if $|\psi\rangle \in \{|1\rangle,|+\rangle \}$, while $C_s=0$ otherwise). For each qubit sent to Alice in a state $|\psi\rangle^{\cal{A}}$, Trent sends to Bob classical bits $C^{\cal{A}}= C(|\psi\rangle^{\cal{A}})$ assigned to this state, and analogously for Bob. 

During the Exchange phase, that might take place much later after the Initialization phase, the clients agree upon the actual contract, decide whether to commit to it or not, and exchange each others' commitments. On the system in the state $|\psi\rangle^{\cal{A}}$, Alice measures either $\hat{A}$ or $\hat{R}$, depending on whether she decides to accept or reject the contract, respectively, and then communicates her result $M^{\cal{A}}$ to Bob. Since the two observables are defined by a pair of mutually unbiased bases, the statistics of measurement results performed on a sequence of qubits each of which is randomly prepared in one of the four states $\{ |0\rangle,|1\rangle,|-\rangle,|+\rangle \}$ will be dramatically different depending on which observable is measured. For instance, when Alice measures $\hat{A}$ on qubits prepared in the Accept basis, the measurement outcomes are perfectly correlated with the corresponding classical bits $C_s^{\cal{A}}$ given to Bob (i.e., $M^{\cal{A}} = C_s^{\cal{A}}$); when she measures $\hat{A}$ on qubits prepared in the Reject basis produces, the results obtained are completely uncorrelated with the corresponding classical bits, and analogously for $\hat{R}$. 

This way, by choosing one of the two measurements performed on a sequence of qubits, Alice produces one of two mutually exclusive sets of measurement outcomes that serve as a signature of her choice (to accept or to reject the contract). By sending the results to Bob, she informs him of her decision by some fixed moment in time $t_0$. The same is done by Bob. 

In the Binding phase, {\em each} party is asked to confront her/his measurement results with the Trent's corresponding classical bits. The perfect correlation between measurement results and the corresponding classical information for qubits prepared in the Accept/Reject basis confirms a client's Accept/Reject choice. Trent declares contract as valid, giving the signed certificates to both clients, if a client, say Alice, accepted the contract, while Bob did not reject it, or vice versa. Note that it is impossible, unless with negligible probability, to produce perfect correlations on both sets of qubits (a direct consequence of the security \cite{qcryptography-security} of the BB84 protocol \cite{bb84}). In this and the next two sections we restrict ourselves to the case of ideal measurements and noiseless quantum channels (from Trent to the clients). The real-case scenario is discussed in the penultimate section of the paper.
%Sec. \ref{sec:fairness-general_measurements}.

Obviously, the Exchange phase as described above suffers from the same problem as any classical protocol not involving information exchange with Trent: upon receiving Alice's results, Bob can stop communication and safely postpone his decision to accept or reject the contract to a later moment in time. Thus, we require that the measurements and the exchange of measurement outcomes between clients happens in steps, (qu)bit by (qu)bit. Nevertheless, if we require perfect correlations between the measuring results and the corresponding classical information for {\it all} qubits distributed to a client, in $1/4$ of the cases an agent will be already after the first step in a position to choose to, with probability one, either bind or reject the contract.

Imagine the following situation. Alice is the first to perform a measurement and send the classical information to Bob. She is the honest party, she wants to accept the contract and thus she measures $\hat{A}$ onto her first qubit. With probability $1/2$, her first qubit will be prepared in one of the states from the Reject basis. By measuring $\hat{A}$ onto one of the $\{ |-\rangle,|+\rangle \}$ states, there is $1/2$ of the probability that her result does not match the value of $C_{s_1}^{\cal{A}}$, and so with probability $1/4$ Alice will not be able to achieve perfect correlations between her measurement outcomes and the corresponding classical data of the qubits prepared in states from the Reject basis. In other words, there is $1/4$ of probability that an honest side will already after the first step of the Exchange phase be unable to reject the contract, even before the other side performed any measurement. In such a situation, Bob can safely stop communication and postpone his choice until later moment in time. 

Thus, we require that in order to accept/reject the contract, a client has to establish perfect correlations on $\alpha N_{A/R}$ qubits prepared in the Accept/Reject basis (with the total number of qubits $N=N_A+N_R$), where $1/2 < \alpha < 1$. We call $\alpha$ the {\it acceptance ratio} (note that, for a protocol to be viable, it is necessary that $\alpha > 1/2$). In this scenario, a client is allowed to obtain $(1-\alpha)N_{A/R}$ wrong results for the states prepared in the basis of her/his choice (Accept or Reject). If $\alpha$ is sufficiently large, this eliminates the possibility of obtaining good enough correlations for both groups of qubits, this way eliminating the above mentioned unfair situation.

\begin{figure}[t]
\centering
\includegraphics[width=8.0cm,height=6.5cm,angle=0]{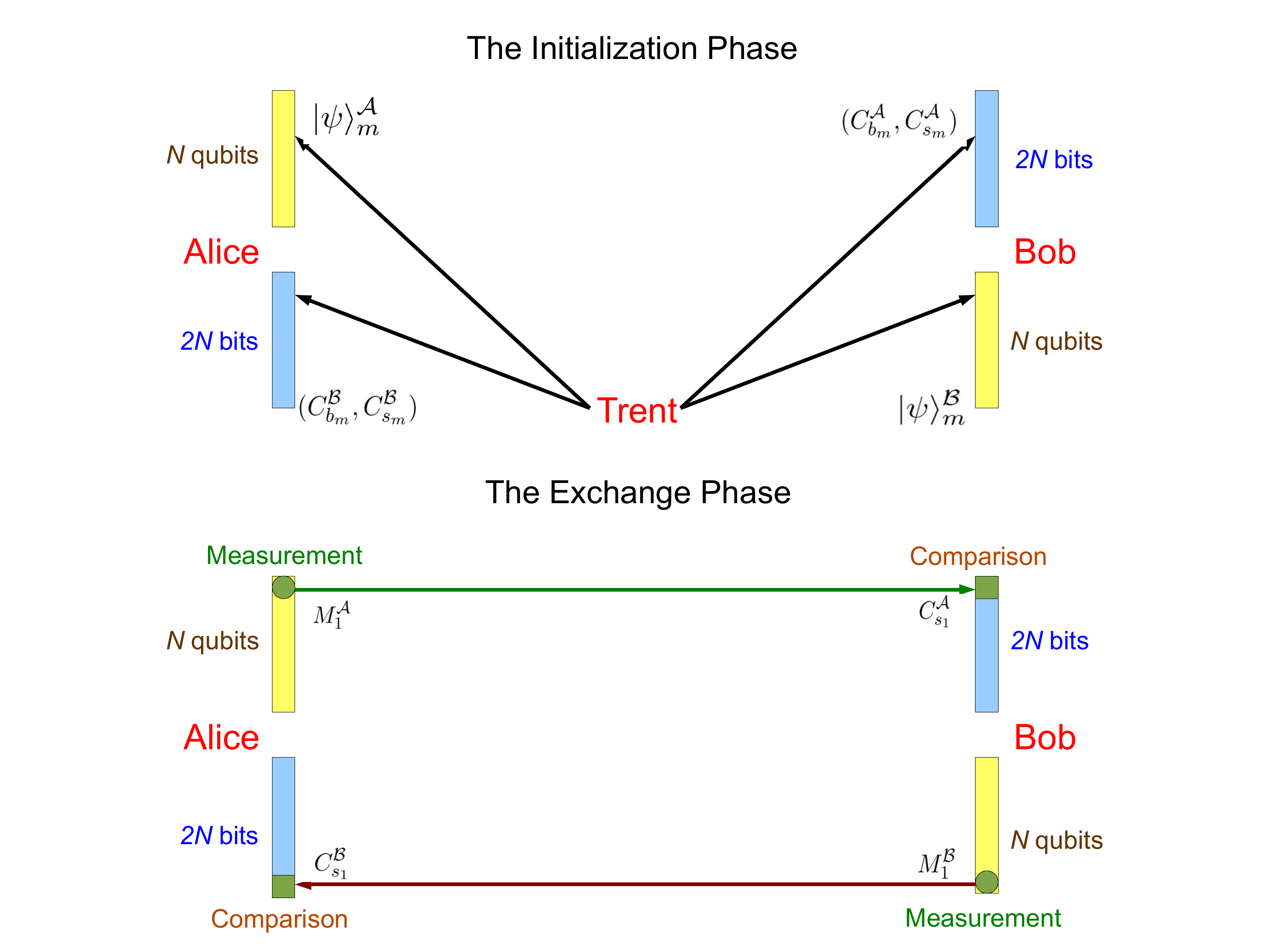}
\caption{(color online) The Initialization and The Exchange Phase.} \label{Initialization and Exchange Figure}
\end{figure}

In the rest of this section, we present a formal description of our protocol. Our protocol is optimistic and as such it is divided into three phases: the Initialization, the Exchange (see Fig. \ref{Initialization and Exchange Figure}) and the Binding phase. During the Exchange phase agents exchange their measurement results. If both clients are honest and perform measurements according to the protocol (measure the Accept observable), the Exchange phase will end up with both clients having their probabilities to bind the contract {\it exponentially} (in number of qubits) close to one: the protocol is viable and optimistic (clients do not need to contact Trent as they already know the answer). If a client, say Bob, is dishonest and performs measurements other than that prescribed by the protocol (or just guesses the outcomes), he will unavoidably obtain wrong outcomes for some of the qubits from the Accept basis. If Alice detects a wrong result (Bob's {\em cheating}), she interrupts the exchange and proceeds to the Binding phase. In a realistic case of measurement errors, Alice will have to set a threshold for the allowed number of  wrong results below which she continues with the exchange. We discuss it at the end of this paper.

%______________________________________________________________________ FIGURE

%________________________________________________________________________________

{\bf The Initialization Phase:} {\it Trent produces $N$ pairs of qubits in states $(|\psi\rangle_m^{\cal{A}},|\psi\rangle_m^{\cal{B}})$ with the corresponding classical description $(C_m^{\cal{A}},C_m^{\cal{B}})=((C_{b_m}^{\cal{A}},C_{s_m}^{\cal{A}}),(C_{b_m}^{\cal{B}},C_{s_m}^{\cal{B}}))$, with $m\in \{1,\ldots N \}$. The rule of assigning the classical data to the corresponding qubit states is the following: $C_{b_m}^{\cal{A/B}}=1$ if $|\psi\rangle_m^{\cal{A/B}} \in \mathcal{B}_A$, while $C_{b_m}^{\cal{A/B}}=0$ otherwise; $C_{s_m}^{\cal{A/B}}=1$ if $|\psi\rangle_m^{\cal{A/B}} \in \{|1\rangle,|+\rangle \}$, while $C_{s_m}^{\cal{A/B}}=0$ otherwise. Each qubit state is randomly chosen from the set $\{ |0\rangle,|1\rangle,|-\rangle,|+\rangle \}$. Trent distributes to Alice $N$ qubits $|\psi\rangle_m^{\cal{A}}$ and $2N$ classical bits $C_m^{\cal{B}}$, and analogously for Bob, keeping the copy of the classical data to himself. He assigns a unique identifier (number) to all data so that it can be linked in the Exchange phase to a specific contract.}

{\bf The Exchange Phase:} {\it Alice and Bob agree on a contract and exchange signed messages containing the contract, the identifier of qubits sequence they want to use, and some previously arranged moment in time $t_0$ giving time restriction to finish the Exchange phase. (This does not bind them to the contract!) Alice and Bob perform measurements on their qubits and exchange the measurement results with each other. Without the loss of generality, we assume Alice is the first to start communication. She measures an observable of her choice ($\hat{A}$ or $\hat{R}$) on the state $|\psi\rangle_1^{\cal{A}}$, obtaining the result $M_1^{\cal{A}} \in \{ 0,1 \}$, and sends it to Bob. Bob compares $M_1^{\cal{A}}$ with $C_{s_1}^{\cal{A}}$. If the values are different, Alice measured her qubit in the basis corresponding to} $(1+C_{b_1}^{\cal{A}}) \ \mbox{mod} \ 2$. {\it Otherwise, the comparison is inconclusive. Next, Bob repeats the procedure described for Alice. The rest of the exchange consists in repeating the above procedure for the states $(|\psi\rangle_m^{\cal{A}},|\psi\rangle_m^{\cal{B}})$ with $m\in \{2,\ldots N \}$. If a client, say Alice, does not obtain a result from Bob until $t_0$ or receives for a qubit from the Accept basis a result different from the corresponding classical data ($C_{b_m}^{\cal{B}}=1 \wedge M_m^{\cal{B}}\neq C_{b_m}^{\cal{B}}$), she immediately proceeds to the Binding phase.}

{\bf The Binding Phase:} {\it At the beginning of the Binding phase Trent chooses $\alpha \in (1/2,1)$ randomly and independently, according to a publicly known probability distribution $p(\alpha)$. Without the loss of generality, we assume that Alice contacts Trent to decide validity of the contract. She decides according to her preference whether she want to bind or to reject the contract. In the former case she measures all unmeasured qubits in the Accept basis, in the latter in the Reject base. Both parties then report Trent for each respective qubit whether they measured it in the Accept or Reject basis, and submit respective measurement outcomes. Trent verifies whether their measurement outcomes correspond to their claims. If there is a mismatch in measurements of, say Bob, he is declared as cheater and Trent considers only Alice's measurement outcomes. Let $N_A^{\cal{A}}$ ($N_R^{\cal{A}}$) denote the number of Alice's qubits prepared in the Accept (Reject) basis, and analogously $N^{\cal{B}}_A$ and $N_R^{\cal{B}}$ for Bob. The contract is declared as valid if Alice presents at least $\alpha N_A^{\cal{A}}$ accept results and Bob presents less than $\alpha N_R^{\cal{B}}$ reject results, or when Bob presents at least $\alpha N_A^{\cal{B}}$ accept results and Alice presents less than $\alpha N_R^{\cal{A}}$ reject results. In case a client, say Bob, supplied incorrect measurement outcomes (see above), Trent declares the contract to be valid if Alice presents at least $\alpha N_A^{\cal{A}}$ accept results. In all other cases the contract is declared as invalid.}

%%%%%%%%%%%%%%%%%%%%%%%%%%%%%%%%%%%%%%%%%%%%%%%%%%%%%%%%%%%%%%%%%%%%%%%%%%%%%%%%%%%%%%%%%%%%%%%%%%%
\section{III. Fairness conditions}
\label{sec:fairness-conditions}
%%%%%%%%%%%%%%%%%%%%%%%%%%%%%%%%%%%%%%%%%%%%%%%%%%%%%%%%%%%%%%%%%%%%%%%%%%%%%%%%%%%%%%%%%%%%%%%%%%%

As noted in the previous section, our protocol is (probabilistically) viable: if both clients are honest, the probability to bind the contract is exponentially close to one. 
Therefore, it is also optimistic: honest clients do not need to contact Trent in order to obtain the verdict that, with exponentially high probability, they already know. In case a client, say Bob, is not honest, i.e. is not measuring the Accept observable in every step of the protocol, we say that he is cheating. Any cheating strategy will inevitably have a non-zero probability of producing a wrong result on qubits from the Accept basis, thus allowing Alice to detect Bob's cheating and move on to the Binding phase. 

In case the Exchange phase is terminated due to cheating detected, all we can predict is the probability that Trent declares the contract as valid. This probability depends on the moment when the exchange was aborted as well as on actions of both parties (before and after the exchange was terminated). The preferences of the signing parties may change (due to commodity price changes, etc.) before it is possible to reach Trent. We say that parties are symmetric, if the probability that Trent declares the contract as valid is (almost) the same regardless whether honest Alice wants to bind the contract and Bob wants to reject it, or vice versa. Note that we do not care about trivial cases when both want to reject or both to bind the contract.

This notion of symmetry is close to the weak coin tossing problem \cite{Mochon-Quant_weak_coin-:2004}: if both Alice and Bob want the same outcome ($0$ or $1$), there is no need to guarantee unbiased coin toss. On the other hand, it is vital to assure as little bias as possible, if their preferences are contradictory.

In this section, we present our formal fairness conditions that allow for such symmetric positions of the clients. In the following, we assume Alice to be honest and Bob dishonest client.

The {\em probability to bind the contract}, $P_{bind}$, is a probability that during the Binding phase Trent, upon receiving classical information from both Alice and Bob (i.e., their measurement results), declares the contract as valid and hands in to both Alice and Bob signed (i.e., validated) copies of the contract.

The probability to bind is a function of step $m$, after which the communication is broken (typically, if cheating is noticed, or a connection between the clients broken). It is also a function of clients' strategies. In this section, we will consider the case of one-particle measurements only. As shown in section V., the general case of multi-particle measurements, acting on at most $L$ qubits, can be reduced to the case of one-qubit measurements, by considering the blocks of $L$ qubits as units and then applying the same reasoning to the sequence of $N/L$ such units. The case of global measurements on all $N$ qubits is discussed at the end of Section V.

Restricting ourselves to one-qubit measurements, for a fixed parameter $\alpha$, the probability to bind is a function of clients' measurements before the step $m$, $X^{\cal{A}}$ for Alice, $X^{\cal{B}}$ for Bob, and their behavior after the step $m$, $Y^{\cal{A}}$ for Alice, $Y^{\cal{B}}$ for Bob. Thus,
\begin{equation}
P_{bind} = P_{bind} (m, \alpha,  (X^{\cal{A}}, X^{\cal{B}}),  (Y^{\cal{A}}, Y^{\cal{B}})).
\end{equation}

%\footnote{For each particular instance, all the relevant probabilities are functions of measurement outcomes. The probabilities that we consider are averaged over all possible outcomes.}

Alice is an honest client, so her strategy is $X^{\cal{A}} = \hat{A}^{\otimes m}$. In Section V. we show that, when measured on qubits prepared in one of the states from the Accept and the Reject bases, general one-qubit measurements produce probability distributions equivalent to those obtained when only the Accept and the Reject observables are measured. Therefore, Bob's strategy, in which he measures the $\hat{A}$ observable $(m - \delta m)$ times, while $\hat{R}$ observable is measured $\delta m$ times (the order is irrelevant), is $X^{\cal{B}} = \hat{A}^{\otimes (m - \delta m)} \hat{R}^{\otimes (\delta m)}$. Thus, we can write $X^{\cal{B}} = X^{\cal{B}} (m, \delta m)$.

The only two relevant strategies after step $m$ are those in which one client wants to bind the contract and the other does not want to bind the contract. The one who wants it bound measures the Accept observable, $Y = \hat{A}^{\otimes (N-m)}$. The other, who does not want it bound, measures the Reject observable, $Y = \hat{R}^{\otimes (N-m)}$. If Alice wants the contract bound, $Y^{\cal{A}} = \hat{A}^{\otimes (N-m)}$, and Bob doesn't, $Y^{\cal{B}} = \hat{R}^{\otimes (N-m)}$, we call the corresponding probability Alice's probability to bind the contract, $P^{\cal{A}}_{bind} (m, \alpha, X^{\cal{A}}, X^{\cal{B}})$, and analogously for Bob (note that in this notation, clients' strategies after step $m$ are not explicitly written). We will calculate clients' probabilities to bind the contract in the next section. Here, we only note that we do not discuss cases when after step $m$ both clients want the same: if the protocol is fair for the cases when clients' wishes are opposite (a conservative assumption), it will be when they wish the same.

Averaging over all possible Bob's strategies $X^{\cal{B}}$ (Alice is honest, so her strategy is known), we obtain
\begin{equation}
P_{bind}^{\cal{A}} (m, \alpha) = \sum_{X^{\cal{B}}} p(X^{\cal{B}}) P_{bind}^{\cal{A}} (m, \alpha, X^{\cal{A}}, X^{\cal{B}}),
\end{equation}
and analogously for $P^{\cal{B}}_{bind} (m, \alpha)$. Here, $p(X^{\cal{B}})$ is the probability that Bob chooses the particular strategy $X^{\cal{B}} = X^{\cal{B}} (m, \delta m)$, and is given by the probability $p_w(\delta m) = 1-(3/4)^{\delta m}$ to obtain a wrong result, when measuring the Reject observable $\delta m$ times (obviously, wrong results are, in the case of ideal measurements with no errors, possible only on qubits from the Accept basis). 

For our protocol to be fair, we require that at each step $m$ of the Exchange phase, the difference between the agents' (Alice and Bob) probabilities to bind the contract can be made arbitrarily small: for any given $\varepsilon$, 
\begin{equation}
|P^{\cal{B}}_{bind}(m, \alpha) - P^{\cal{A}}_{bind}(m, \alpha) | < \varepsilon.
\end{equation}
%Note that the very choice of the strategy depends on the measurement outcomes: Bob is cheating as long as his cheating is not detected -- as long as his measurement outcomes allow him to continue the cheating strategy. As noted, the averaging should be also done over all possible measurement outcomes, as well as over all possible distributions of states from the Accept and the Reject bases, which for simplicity is taken to be implicit. 

In order to make our protocol even more symmetric, we introduce {\em the probability to cheat} of a dishonest client (Bob). It is the product of Bob's probability to bind, and the probability that Alice will not bind the contract:
\begin{eqnarray}
P_{ch}^{\cal{B}} (m, \alpha, X^{\cal{A}}, X^{\cal{B}}) &=& P^{\cal{B}}_{bind} (m, \alpha, X^{\cal{A}}, X^{\cal{B}}) \\ 
&& \times [1 - P^{\cal{A}}_{bind} (m, \alpha, X^{\cal{A}}, X^{\cal{B}})]. \nonumber
\end{eqnarray}

After the averaging over Bob's strategies $X^{\cal{B}} (m)$ probability to cheat of a dishonest client Bob is:
\begin{equation}
P_{ch}^{\cal{B}} (m, \alpha) = \sum_{X^{\cal{B}}} p(X^{\cal{B}}) P_{ch}^{\cal{B}} (m, \alpha, X^{\cal{A}}, X^{\cal{B}}).
\end{equation}

Our second fairness requirement is that a dishonest client's (Bob's) probability to cheat $P_{ch}^{\cal{B}} (m, \alpha)$ is also negligible\footnote{Obviously, if the second fairness requirement is satisfied, the first if satisfied as well: having negligible probability to cheat is stronger condition than the symmetry between the client's probabilities to bind the contract.}. Note that in our protocol the Binding phase requires {\em both} clients to confront their measurement results, {\em both} obtaining the {\em same} verdict by Trent at the end. Therefore, although the probability to cheat is the product of two probabilities, the probability that Bob can bind the contract and the probability that Alice cannot, it is not itself a probability of an event (Bob's probability to bind is obtained under the assumption that after step $m$ he measures $\hat{A}$ and Alice $\hat{R}$; to obtain Alice's probability to bind, we assume that after step $m$ she is the one to measure $\hat{A}$ while Bob measures $\hat{R}$). Yet, it can serve as a measure of protocol's fairness as it quantifies agent's freedom to choose between binding and refusing the contract later in time. We discuss this in more detail at the end of this section. At the end of the paper, we show that it is straightforward to design a protocol in which it is not necessary that both clients are (simultaneously) present during the binding phase. In this scenario, the corresponding probability to cheat becomes the joint probability that one client can bind the contract, while the other cannot.

Unfortunately, although the first fairness criterion is satisfied for the above protocol, its probability to cheat can be as high as $1/4$: in case both clients measure the Accept observable, both clients' probabilities to bind are monotonic functions, such that for big enough $N$ there exists $m_s$ for which $P^{\cal{A}}_{bind}(m_s, \alpha) \approx P^{\cal{B}}_{bind}(m_s, \alpha) \approx 1/2$. But if $\alpha$, determined by Trent by sampling a random variable described by a publicly known probability distribution $p(\alpha)$, is itself unknown to the clients, and determined only later during the Binding phase, the expected probability to cheat $\bar{P}_{ch}(m)$ becomes negligible, for big enough $N$. Thus, our second fairness condition finally reads as: for any given $\varepsilon$,
\begin{equation}
 \! \bar{P}^{\cal{B}}_{ch}(m) \! =  \! \int p(\alpha) P_{bind}^{\cal{B}}(m,\alpha)[1-P_{bind}^{\cal{A}}(m,\alpha)]{\mathbf d\alpha} < \varepsilon.
\end{equation}
The coefficient $\alpha$ is sampled randomly by Trent to achieve stronger security requirements. This assures symmetric position of honest and cheating participant even before Trent is contacted during the Binding phase: if agents are temporarily unable to contact Trent, cheaters should not profit from this in a significant way.%, e.g. due to denial of service attack,

A client, say Bob, may be willing to take the risk and stop the protocol prematurely during the Exchange phase, provided such a situation can assure him some reasonable position. Consider a contract where Alice buys orange juice from Bob for $X$ units per litter. According to the market expectation, with probability $p$  the price should increase and with probability $(1-p)$ decrease. When the price goes up to $X'>X$, Alice wants to enforce the contract, since otherwise she should buy juice for higher price. Bob wants the contract to be canceled to sell the juice for higher price. In case the price drops, the situation is symmetric.

%p \!\! \in \!\! [0,1]

Bob may be willing to take the risk parameterized by $\delta$ in the following sense. The joint probability that the price drops and he will be able to enforce the contract is at least $\delta$ as well as the joint probability that price increases and Alice won't be able to enforce the contract. The latter gives him protection from financial loses, while the former allows him to spare some money. This is formalized as 
\begin{equation}
	(\exists\, 0 \! \le \! p \! \le \! 1\ \! ) \left[p(1-P_{bind}^{\cal{A}})\geq\delta\,\wedge\ (1-p)P_{bind}^{\cal{B}}\geq\delta\right].
\end{equation} 
Thus, to prevent reasonability of Bob's cheating we require that 
\begin{equation}
(\forall\, 0 \!\! \le \! p \! \le \!\! 1\ \!\! ) \left[p(1-P_{bind}^{\cal{A}})\le\delta\,\vee\ (1-p)P_{bind}^{\cal{B}}\le\delta\right].
\end{equation}
Let us denote $Y\stackrel{def}{=}P_{bind}^{\cal{B}}(m,\alpha)[1-P_{bind}^{\cal{A}}(m,\alpha)]$ the random variable parameterized by $\alpha$. The expected value 
\begin{equation}
	E(Y)=\int p(\alpha) P_{bind}^{\cal{B}}(m,\alpha)[1-P_{bind}^{\cal{A}}(m,\alpha)]{\mathbf d\alpha} %\equiv \bar{P}_{ch}(m)
\end{equation} 
is nothing but the {\it expected probability to cheat} $\bar{P}_{ch}(m)$ (note that due to $P^{\cal{B}}_{bind}(m, \alpha) \approx P^{\cal{A}}_{bind}(m, \alpha)$, we have $\bar{P}_{ch}(m) \equiv \bar{P}^{\cal{B}}_{ch}(m) \approx \bar{P}^{\cal{A}}_{ch}(m)$ ). Using the above fairness criterion $\bar{P}_{ch}(m) < \varepsilon$, Chebyshev inequality, and putting $\delta^3=\varepsilon$, we obtain $\mbox{Prob}_\alpha[Y<\delta+\delta^3]\geq 1-\delta.$ Thus, the probability $\delta$ can be made arbitrarily small with arbitrarily high probability.

%%%%%%%%%%%%%%%%%%%%%%%%%%%%%%%%%%%%%%%%%%%%%%%%%%%%%%%%%%%%%%%%%%%%%%%%%%%%%%%%%%%%%%%%%%%%%%%%%%%
\section{IV. Fairness of the protocol: Ideal case}
\label{sec:fairness-ideal}
%%%%%%%%%%%%%%%%%%%%%%%%%%%%%%%%%%%%%%%%%%%%%%%%%%%%%%%%%%%%%%%%%%%%%%%%%%%%%%%%%%%%%%%%%%%%%%%%%%%

In the following, we show that our protocol is {\em fair}: both  $|P^{\cal{B}}_{bind}(m, \alpha) - P^{\cal{A}}_{bind}(m, \alpha)|$ and $\bar{P}_{ch}(m)$ could be made arbitrarily low. In this section, we assume that only $\hat{A}$ or $\hat{R}$ are measured, and that no measurement errors or qubit state corruption occur. In the next section, we discuss general (one or multi-qubit) observables and real-life scenario of imperfect measurements and noisy channels.

In case Bob is cheating during the Exchange phase, he will be detected after a small number of steps, with probability growing exponentially in the number of qubits measured by Alice. Assume Bob's cheating is detected after Alice measured $m$ qubits. Alice terminated the Exchange phase and participants proceed with the Binding phase, that can be delayed (Trent is offline). Meanwhile, participants are allowed to change their preferences and we would like to examine symmetry of their position. We are interested only in the situation when Alice wants to bind the protocol and Bob wants to reject, and vice versa.

In the former case Alice tries to do her best to bind the contract. This means she measures all unmeasured qubits in the Accept basis and sends her results to Trent. Bob does his best to invalidate the contract, measuring his qubits in the Reject basis.
Note that possible lying about measurement basis on respective qubit is detected with probability growing exponentially in the number of wrongfully reported measurement results, so the number of measurements Bob can lie about is well limited.

%First, note that it is impossible for Bob to reproduce {\em all} classical data corresponding to his qubit states, even when general measurements are allowed, unless with negligible probability (which secures the BB84 protocol \cite{bb84,qcryptography-security} as well). The probability to guess the classical data is continuous in $\alpha$, thus Bob cannot pass {\em both} the commit and the reject test during the binding phase, for a suitable range of $\alpha$. Therefore, he must trick Alice by playing a strategy such that after a certain step $m$ in the results' exchange, with high probability Alice is no longer able to reject the contract while Bob can. Since only $\hat{A}$ or $\hat{R}$ are measured, Bob starts the information exchange by measuring his qubits in the Accept basis, pretending to be an honest side. 

If the cheating was detected and the Exchange phase was terminated after $m$ steps, Alice's probability to bind the contract is, for a given $\alpha$,                                                                                                                                                                                                                                                                                                                                                                                                              given by the following expression:
\begin{equation}
P^{\cal{A}}_{bind} (m, \alpha, X^{\cal{A}}, X^{\cal{B}}) = P^{\cal{A}}_A (m, \alpha, X^{\cal{A}}) [1 - P^{\cal{B}}_R (m, \alpha, X^{\cal{B}})],
\end{equation}
and analogously for Bob. Probability to accept the contract, $P_A$, is the probability that a client who wants the contract bound, thus after step $m$ measuring the Accept observable, will pass the Trent's test during the Binding phase. Probability to reject the contract, $P_R$, is the probability that a client who does not want the contract bound, thus after step $m$ measuring the Reject observable, will invalidate the contract during the Binding phase. Note that under this convention it is redundant to specify clients' strategies $Y$ after the step $m$. Also, clients' probabilities to accept and reject the contract depend only upon their own strategies before the step $m$, and not those of the opponent.

Alice's probability to accept the contract is $P^{\cal{A}}_A (m, \alpha) = 1$: she is an honest client and thus is measuring $\hat{A}$ in every step until the interruption; she wants the contract bound, so continues to measure the Accept observable. Her probability to reject the contract can be written in a simplified form as $P^{\cal{A}}_R (m, \alpha)$, bearing in mind that in this case $X^{\cal{A}} (m) = \hat{A}^{\otimes m}$ and $Y^{\cal{A}} (m) = \hat{R}^{\otimes (N-m)}$.

Out of $m$ steps, a dishonest Bob measures the Accept observable $(m - \delta m)$ times, and the remaining $\delta m$ times the Reject observable. Therefore $X^{\cal{B}} (m, \delta m) = \hat{A}^{\otimes (m - \delta m)} \hat{R}^{\otimes (\delta m)}$. When rejecting the contract, his strategy is $Y^{\cal{B}} (m) = \hat{R}^{\otimes (N-m)}$, and we see that Bob's probability to reject is the same as Alice's probability to reject, had the interruption occurred $\delta m$ steps before, $ P^{\cal{B}}_R (m, \alpha, \delta m) = P^{\cal{A}}_R (m-\delta m, \alpha)$. When measuring the Reject observable, the probability to obtain wrong results on states from the Accept basis, and thus being detected cheating, is exponentially fast approaching to one, $p_w(\delta m) = 1-(3/4)^{\delta m}$. Therefore, for $1/2 < \alpha < 1$ and large enough $N$, $\delta m << (1-\alpha )N$. The expected $\delta m$ is small and since for large enough $N$ all the probabilities are slow functions of $m$, we have $ P^{\cal{B}}_A (m, \alpha, \delta m) \approx P^{\cal{A}}_A (m, \alpha) = 1$, while $P^{\cal{A}}_R (m, \alpha) \approx P^{\cal{A}}_R (m-\delta m, \alpha) = P^{\cal{B}}_R (m, \alpha)$. Therefore, the first fairness condition $P^{\cal{A}}_{bind}(m, \alpha) \approx P^{\cal{B}}_{bind}(m, \alpha)$ is satisfied.

To show that the second fairness condition is also satisfied, we first note that, due to $P^{\cal{A}}_R (m, \alpha) \approx P^{\cal{B}}_R (m, \alpha)$ and $ P^{\cal{B}}_A (m, \alpha) \approx P^{\cal{A}}_A (m, \alpha) = 1$, the two probabilities to cheat are almost the same and can be, for a given $\alpha$, written in terms of single expected probability to reject the contract
\begin{equation}
	\label{ch2}
P_{ch}(m;\alpha)=P_R(m;\alpha)(1-P_R(m;\alpha)).
\end{equation}
The expected probability to reject, for a given $\alpha$, is
\begin{equation}
P_R(m;\alpha) = \sum_{N_R=0}^{N} q(N_R)P_R(m;\alpha, N_R), 
\end{equation}
where $P_R(m;\alpha, N_R)$ is the probability to (be able to) reject the contract (obtain less than $(1-\alpha)N_R$ wrong results on qubits from the Reject basis, measuring the Accept observable on the first $m$ qubits), for a given acceptance ratio $\alpha$ and the number of qubits prepared in the Reject basis $N_R$, keeping for simplicity the $N$ dependence implicit, and $q(N_R) = 2^{-N}\left( \begin{array}{c} N \\ N_R \end{array} \right)$ is the probability to have exactly $N_R$ states from the Reject basis. Note that $P_R(m;\alpha, N_R)$ is calculated under the assumption that the Accept observable is measured during the first $m$ steps, while after that the Reject observable is measured for the remaining $(N-m)$ steps. 

For $m<(1-\alpha)N_R$ there is always a chance to reject the contract, thus $P_R(m;\alpha, N_R) = 1$. Otherwise,
\begin{equation}
	\label{reject}
P_R(m;\alpha, N_R) = \sum_{n=n^\prime}^{m^\prime} P(n \mbox{ in R};m, N_R)P_R(n \mbox{ in R};\alpha, N_R).
\end{equation}
Here, the probability that exactly $n$ out of the first $m$ qubit states are from the Reject basis is given by\footnote{Since each sequence of $N$ qubit states, prepared such that exactly $N_R$ are from  the Reject basis, is equally probable, the probability $P(n \mbox{ in R};m)$ is given by the ratio of the number of sequences whose $n$ out of $N_R$ qubits prepared in the Reject basis are among the first $m$ qubits, while the other $(N_R-n)$ are among the rest of $(N-m)$ qubits, $\left( \begin{array}{c} m \\ n \end{array} \right) \left( \begin{array}{c}N- m \\ N_R-n \end{array} \right)$, and the total number of allowed sequences is $\left( \begin{array}{c} N \\ N_R \end{array} \right)$.} 
\begin{equation}
P(n \mbox{ in R};m, N_R) = \left( \begin{array}{c} m \\ n \end{array} \right) \left( \begin{array}{c}N- m \\ N_R-n \end{array} \right) \left( \begin{array}{c} N \\ N_R \end{array} \right)^{-1},
\end{equation}
while the probability of being able to reject the contract if $n$ qubits are from the Reject basis is given by\footnote{When measuring the Accept observable $\hat{A}$ on a state from the Reject basis each of the two possible results are equally probable. Therefore, each sequence of $n$ such measurement results has the probability $2^{-n}$. If exactly $i$ out of $n$ individual-qubit results do not match those given by Trent to the other client, and if $i<(1-\alpha)N$, then it is still possible to reject the contract. There are $\left( \begin{array}{c} n \\ i \end{array} \right)$ of such measurement results, for given $n$.}
\begin{equation}
P_R(n \mbox{ in R};\alpha, N_R) =  2^{-n} \sum_{i=0}^T \left( \begin{array}{c} n \\ i \end{array} \right),
\end{equation}
where $T=(1-\alpha)N_R-1$ if $n\geq (1-\alpha)N_R$ and $T=n$ otherwise. Due to the constraint of having exactly $N_{A/R}$ qubits from the Accept/Reject basis, the range of the summation for $n$ in equation \eqref{reject} is given by $n^\prime=0$ for $m\leq N_A$ while $n^\prime=m-N_A$ otherwise, and $m^\prime=m$ for $m\leq N_R$ while $m^\prime=N_R$ otherwise.

Finally, the expected probability to cheat, with respect to a given probability distribution $p(\alpha)$ on the segment $I_{\alpha}$, is 
\begin{equation}
\bar{P}_{ch}(m) = \int_{I_{\alpha}}p(\alpha)P_{ch}(m;\alpha){\mathbf d\alpha}.
\end{equation}

\begin{figure}[!thb]
\centering
%\ \\[-0.5cm]
\includegraphics[width=8.5cm,height=6.0cm,angle=0]{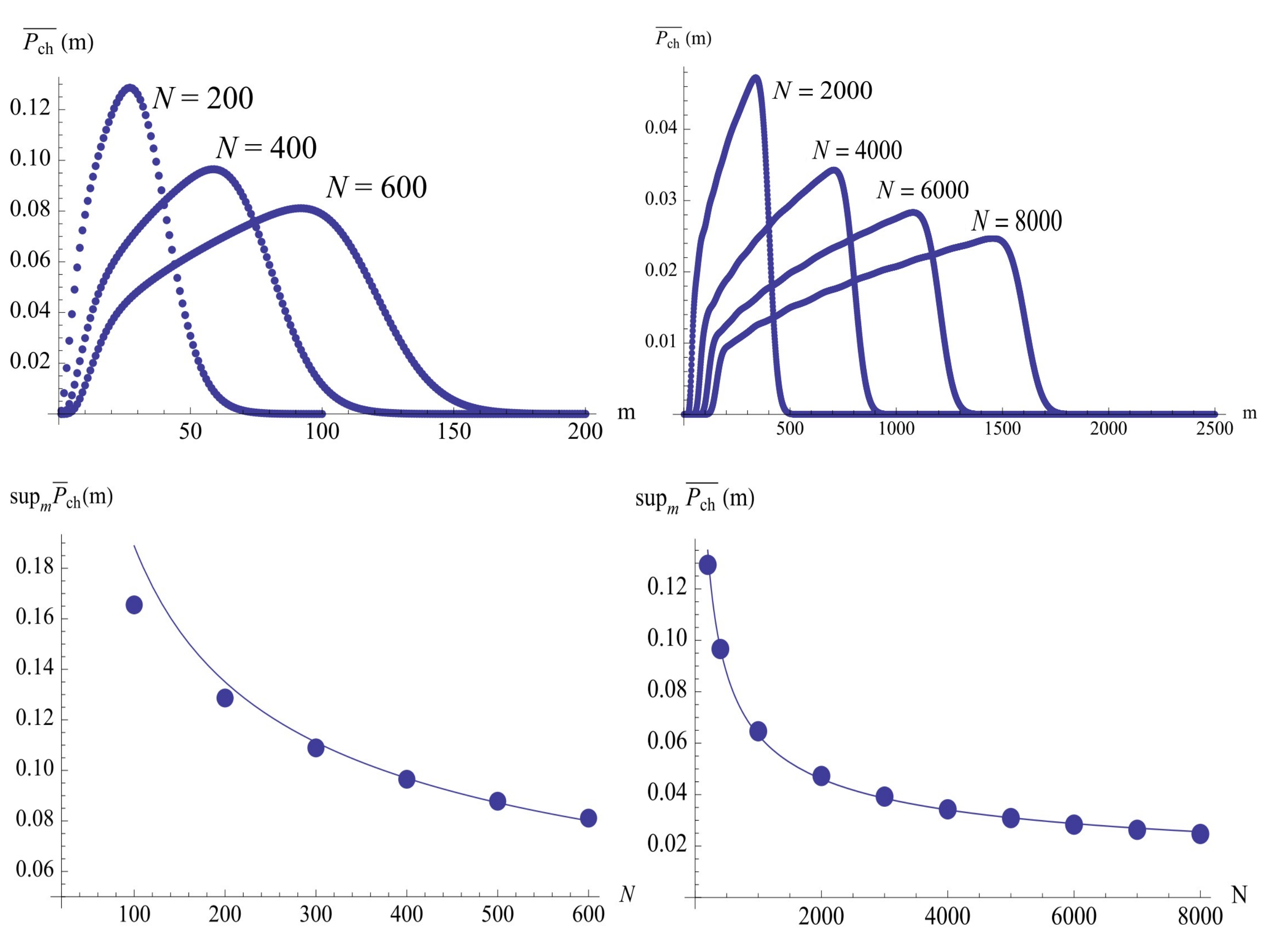}
%\ \\[-0.4cm]
\caption{(color online) The expected probability to cheat $\bar{P}_{ch}(m)$ (upper row) and the maximal expected probability to cheat $\mbox{sup}_m \bar{P}_{ch}(m)$ (lower row) for the uniform $p(\alpha)$ on $I_{\alpha}=[0.9,0.99]$. The plots from the left column represents results for our protocol, while the right ones are for the restricted ``typical'' case of $N_A=N_R$.  Note the scaling behavior $\mbox{sup}_m \bar{P}_{ch}(m) \propto N^{-1/2}$.\ \\[-0.6cm]} \label{Results Figure}
\end{figure}

Using the simplest uniform probability $p(\alpha)=1/I_{\alpha}$ on the segment $I_{\alpha}=[0.9,0.99]$, we numerically evaluated the expected probability to cheat $\bar{P}_{ch}(m)$ for up to $N=600$, while for the ``typical'' case of $N_A=N_R$ we managed to evaluate it up to $N=8000$, see Fig. \ref{Results Figure}. We see that the {\em fairness condition} $\mbox{sup}_m \bar{P}_{ch}(m) << 1$ is satisfied (for $N=600$ we got $\mbox{sup}_m \bar{P}_{ch}(m) = \bar{P}_{ch}(92)= 0.0811$, while for $N_A=N_R$ we have $\mbox{sup}_m \bar{P}_{ch}(m) = \bar{P}_{ch}(1455)= 0.0247$ for $N=8000$), which is explicitly shown on Fig. \ref{Results Figure}. Moreover, the numerical fit gives $\mbox{sup}_m \bar{P}_{ch}(m) \propto N^{-1/2}$ behavior, giving us the scaling of the complexity of the protocol, with respect to the number $N$ of exchanged messages between Alice and Bob.

%______________________________________________________________________ FIGURE

%______________________________________________________________________

%%%%%%%%%%%%%%%%%%%%%%%%%%%%%%%%%%%%%%%%%%%%%%%%%%%%%%%%%%%
\section{V. Fairness of the protocol: General measurements and noise}
\label{sec:fairness-general_measurements}
%%%%%%%%%%%%%%%%%%%%%%%%%%%%%%%%%%%%%%%%%%%%%%%%%%%%%%%%%%%%%

First, we consider only one-qubit orthogonal measurements. Since Alice is an honest client, $X^{\cal{A}} = \hat{A}^{\otimes m}$, we have that $P^{\cal{A}}_A (m, \alpha, X^{\cal{A}}) = 1$ and $P^{\cal{A}}_R (m, \alpha, X^{\cal{A}}) = P_R(m, \alpha)$.

Bob is a dishonest client and his strategy $X^{\cal{B}}$ consists of measuring $(m-k)$ times the Accept observable \!$\hat{A}$ and $k$ times observable $\hat{K}= 0\cdot |m\rangle\langle m| + 1\cdot |m^\bot\rangle\langle m^\bot|$, where 
\begin{equation}
\!\! |m\rangle \! = \! \cos\frac{\theta}{2}|0\rangle + e^{\varphi}\!\sin\frac{\theta}{2}|1\rangle \! = \! \cos\frac{\theta^\prime}{2}|-\rangle  +  e^{\varphi^\prime}\!\!\sin\frac{\theta^\prime}{2}|+\rangle. 
\end{equation} 
Let $m = m_a + m_r$, where $m_a$ is the number of the Accept and $m_r$ the number of the Reject qubits among the first $m$ qubits, and let $k_a$ be the number of measurements of $\hat{K}$ on $m_a$ qubits from the Accept basis, and analogously for $k_r$. Bob's measurements of $\hat{K}$ are equivalent to\footnote{First, we consider measurements of $\hat{K}$ on qubits whose states are from the Accept basis. Let $p_k(0|0)$ be the conditional probability to obtain result $0$ when measuring an observable $\hat{K}$ on the state $|0\rangle$, and analogously for other conditional probabilities. Notice that, when measuring an arbitrary one-qubit orthogonal observable $\hat{K}$, one has $p(0|0)=p(1|1)=s_a$: the probabilities to obtain the right result on the states from the Accept basis are the same. The two probabilities are also equal when one measures $\hat{A}$ with frequency (i.e., probability) $q_a$ and $\hat{R}$ with frequency $1-q_a$. The probability to obtain the right result on the states from the Accept basis is then $S_a= q_a+(1-q_a)/2=(1+q_a)/2$: when measuring $\hat{A}$ (with probability $q_a$) one obtains the right result with probability $1$, while when measuring $\hat{R}$ (with probability $1-q_a$) one obtains the right result with probability $1/2$. In order for those two measurements to be equivalent, it is necessary that $s_a=S_a=(1+q_a)/2$. In the case of the orthogonal observable $\hat{K}$, we have $s_a=\cos^2(\theta/2)$, and thus $q_a=\cos\theta$. The same argument applies for the case of measurements performed on the states from the reject basis.} (for simplicity, we omit writing the $\alpha$ and $N_R$ dependences): 
\begin{itemize}
	\item $q_a\cdot k_a$ measurements of $\hat{A}$ and $\delta m_a = [1-q_a]\cdot k_a$ measurements of $\hat{R}$ on qubits from the Accept basis, where $q_a=\cos\theta$. Thus, the probability to notice cheating after $k_a$ measurements of $\hat{K}$ is $\tilde{p}_w(\delta m_a)=\tilde{p}_w([1-q_a] k_a)= 1-(1/2)^{[1-q_a] k_a}$ and Bob's probability to accept the contract is $P^{\cal{B}}_A (m, \alpha, X^{\cal{B}}) \equiv P_A (m, \alpha; m_a, \delta m_a)$;
	\item $q_r\cdot k_r$ measurements of $\hat{A}$ and $\delta m_r = [1-q_r]\cdot k_r$ measurements of $\hat{R}$ on qubits from the Reject basis, where $q_r=\cos\theta^\prime$. Thus, Bob's measurements are equivalent to $(m_r - \delta m_r)$ measurements of $\hat{A}$ on qubits from Reject basis and his probability to reject is $P^{\cal{B}}_R(m, \alpha, X^{\cal{B}}) \leq P_R(m-\delta m_r,\alpha)$.
\end{itemize}
Since $\tilde{p}_w(\delta m_a) = 1-(1/2)^{[1-q_a] k_a}$, either $k_a$ is small, or $q_a \approx 1$. In case of $q_a \approx 1$, we have that $q_r \approx 0$: the observable $\hat{K}$ is close to $\hat{A}$ and Bob's strategy is close to that of an honest client. In case, $k_a$ is small, we have that $\delta m_a$ is also small and Bob's probability to accept the contract is close to one, $P^{\cal{B}}_A (m, \alpha, X^{\cal{B}}) = P_A (m, \alpha; m_a, \delta m_a) \approx 1$. Moreover, since for typical cases $k_a \approx k_r$, we have that typically Bob's probability to reject the contract is $P^{\cal{B}}_R(m, \alpha, X^{\cal{B}}) \approx P_R(m,\alpha)$. Therefore, the corrected probability to cheat, averaged over all possible distribution of states (from the Accept and the Reject bases) and all possible strategies of the clients, will not be considerably altered and the protocol would still be fair, even if arbitrary number of observables $\hat{K}_i$ is allowed. Note that, according to the above and the analysis from the previous section, the complexity of the protocol, for the case of general one-qubit orthogonal measurements, scales as $N \propto \varepsilon^{-2}$, where $N$ is the number of exchanged messages and $\varepsilon$ a given upper bound for the probability to cheat.

The above argument could be generalized for joint $L$-qubit measurements, if $L \propto N^t$ and $t<1$: for every joint observable (or general POVM) 
$\hat{O}_L \neq \otimes_{i=1}^L\hat{A}_i$ 
there is a non-zero probability $q_L$ that at least one wrong result will be obtained on the accept qubits, which scales as $q_L^k$, $k$ being 
the number of $\hat{O}_L$ measurements (note that for ideal measurements, Alice will notice cheating as soon as she receives the {\em first} wrong result from Bob). Thus, in case of 
performing joint  
measurements on at most $L$ qubits, for $N/L >> 1$, the protocol would still be fair. 
%Note that for a fixed $L$, according to the above and the analysis from the previous section, the complexity of the protocol scales as $N \propto \varepsilon^{-2}$, where $N$ is the number of exchanged messages and $\varepsilon$ a given upper bound for the probability to cheat. 

In case $L \propto N$ the fairness of our protocol could be seen as a consequence of the unconditional security of the BB84 protocol in the presence of noisy channel and imperfect sources and detectors \cite{qcryptography-security}: if we interpret $(1- \alpha)$ as the error rate due to the noise (see the paragraph below), then in order to successfully cheat in our contract signing protocol, an agent would have to be correct on {\em both} $\alpha N_A$ qubits from the Accept basis and on $\alpha N_R$ qubits from the Reject bases, which in turn contradicts the unconditional security of the BB84 protocol. Namely, in order to pass the test by Alice and Bob and thus learn the secret key, Eve has to know the states of qubits from both mutually unbiased bases. In the presence of errors, she needs to be correct only on the $\alpha$ fraction of qubits. Since quantum cryptography is unconditionally secure, that is not possible, unless with exponentially small probability. Thus, our quantum contract signing protocol is fair. Note that, although the errors due to imperfect technology are inevitable, it is possible, at least in principle, to make them arbitrary low. Therefore, if for some error rate $(1- \alpha)$ it is not possible, unless with negligible probability, to have the right results on $\alpha N_{A/R}$ qubits from both the Accept and the Reject bases, then for some better equipment the fairness condition would be satisfied.
%which is, due to continuity in $\alpha$ of the probability to guess the classical data, for a suitable range of $\alpha$ impossible unless with negligible probability.

In the case of measurement errors and noisy channels, one must introduce the error tolerance $\eta = M_w/M$, where $M_w=\langle m_w\rangle\equiv\eta M$ is the expected number of wrong results obtained in measuring an observable on $M$ qubits prepared in states from the observable's eigenbasis. Coefficient $\eta$ gives the ratio of unavoidably produced wrong results: to detect cheating would then mean to obtain more than expected, according to $\eta$, wrong results. For $\eta < (1- \alpha)$ and big enough $N$, our protocol would therefore still be fair. 

Our protocol can be modified to use only two non-orthogonal pure states, just like the B92 cryptographic protocol \cite{b92} is modified after the BB84. Obviously, the probabilities determining the features and complexity of the protocol would be quantitatively different, but the protocol would clearly still be fair.

We note that it is straightforward to design a protocol in which a single client can contact Trent to obtain a signed contract. In this case, Trent sends to a client, say Alice, classical information about only a half of, randomly chosen, Bob's qubit states. This information is used by Alice to check Bob's measurements: whether he is measuring the Accept observable, or not. If so, he is measuring the Accept observable on the rest of his qubits as well (he does not know for which qubits Alice has the classical information and for which not). Thus, the results Bob provided her for the rest of his qubits is used by Trent to verify Alice's data: instead of Bob appearing with his results, Alice provides those given to her by Bob. Note that in this case, unlike before, the corresponding average probability to cheat is probability of a real event: joint probability that an agent cannot bind the contract, while the other can.

This version of the protocol is also suited for the use of entangled pairs, instead of qubits in definitive pure states. In this case, Trent produces $4N$ pairs of maximally entangled qubits in the singlet state $|\psi^- \rangle = (|01\rangle - |10\rangle )/\sqrt{2}$. Half of each pair of the first $2N$ entangled pairs are distributed to Alice, while the other half is kept with Trent. Each half of the other $2N$ pairs is sent to Bob. Then, Trent randomly measures either of the two observables (Accept or Reject) on each of $4N$ qubits left with him. Random $N$ results obtained from measuring the first $2N$ qubits are sends to Bob (together with the classical information of their positions within the first $2N$ pairs), and analogously with the other $2N$ results ($N$ of which are sent to Alice). We see that this situation is equivalent to the above, when instead of entangled, Trent sends qubits in pure states. 

Finally, we note that it is simple to generalize our protocol for the case of arbitrary number of parties. In case of $n$ parties, a classical information about the preparation of each of $N$ qubits sent to, say the first client, is divided into $N/(n-1)$ sets $S_i$, with $i = 2, \ldots n$, such that each set $S_i$ is sent to the $i$-th client. The protocol is executed in $N$ rounds, such that in the $i$-th step of each round, the $i$-th client is performing a single-qubit measurement and publicly announces the result to all of the remaining $(n-1)$ clients. To declare contract valid, t binding all of the $n$ clients, the requirements analogous to the ones for the two-party protocol must be satisfied between each of the $\left( \begin{array}{c} n \\ 2 \end{array} \right)$ pairs of clients. This, as well as the above mentioned generalizations of our protocol, will be discussed in more detail elsewhere.

%%%%%%%%%%%%%%%%%%%%%%%%%%%%%%%%%%%%%%%%%%%%%%%%%%%%%%%%%%%%%%%%%%%%%%%%%%%%%%%%%%%%%%%%%%%%%%%%%%%
\section{VI. Conclusions}
\label{conclusions}
%%%%%%%%%%%%%%%%%%%%%%%%%%%%%%%%%%%%%%%%%%%%%%%%%%%%%%%%%%%%%%%%%%%%%%%%%%%%%%%%%%%%%%%%%%%%%%%%%%

We have presented a fair and optimistic quantum protocol for signing contracts that does not require the exchange of information with the trusted party during the Exchange phase. Unlike the classical proposals, its fairness is based on the laws of physics rather than on sending signed messages, that are only computationally secure. Thus, no keys are generated during the Exchange phase and the protocol is abuse-free. For single-qubit orthogonal measurements, the complexity of the protocol scales as $N \propto \varepsilon^{-2}$, where $N$ is the number of exchanged messages and $\varepsilon$ given threshold for the probability to cheat. Analogously to the previous proposal of a quantum contract signing protocol \cite{qsig-previous}, the present one could also be performed using either entangled pairs or two, instead of four pure states, but it does not require tamper-proof devices nor is based on the effects of decoherence (which are necessary for achieving \cite{qsig-previous}). Also, it is simple to generalize it to involve many clients.  Our protocol can easily be altered such that only one client is enough to present her/his results in order to bind the contract and obtain the document declaring the contract as valid. In this case, the probability to cheat becomes a joint probability that a client can bind a contract, while the other cannot. The fairness condition of a negligible probability to cheat is in this case slightly weaker than the one of \cite{ben:90}. Also, in our quantum protocol the trusted party is slightly more involved than in the classical protocol presented in \cite{rabin:83}.  Nevertheless, being based on the laws of physics, our protocol has other security advantages over the classical counterparts. Moreover, it can be used for designing other quantum information and security protocols. Indeed, our approach was used to design the protocols of simultaneous dense coding and teleportation between spatially distant clients \cite{daowen:11}. Finally, our protocol could be easily performed with the current technology used in quantum cryptography.

The possible future lines of research include a detailed study of the effects of noise and multi-particle measurements on the fairness of the protocol and its complexity. Today's quantum technology is still at its infancy and experimental realizations and technological applications of quantum information protocols are relatively rare and not always as reliable as classical alternatives. Even the security of quantum cryptography, that is proven to be unconditionally secure under the effects of noise, is still very much dependent on the technology used. Nevertheless, the development of quantum technology has already proven to be remarkable and beyond many of early expectations, and will for sure continue to develop. Another possible line of future research could be a further study of alternative versions of quantum contract signing protocols with possibly better properties and designing novel quantum information, security and computation protocols involving spatially distant clients that require timely decisions.

\begin{acknowledgments}
NP and PM thank the project of SQIG at IT, funded by FCT and EU FEDER projects QSec PTDC/EIA/67661/2006 and QuantPrivTel PTDC/EEA-TEL/103402/2008, IT Project QuantTel, and Network of Excellence, Euro-NF. %	and the SQIG LAP initiative.
The authors acknowledge discussions with V. Bo\v{z}in and \v{C}. Brukner.
\end{acknowledgments}

%\bibliography{qsecPRL}
%\bibliographystyle{plain}
%\bibliographystyle{splncs}

\end{document}